\documentclass[12pt]{article}
\usepackage{amsmath}
\usepackage{amssymb}
\usepackage{epsfig}
\usepackage{cite}
\usepackage{color,colordvi}

\begin{document}
\vspace{0.01cm}
\begin{center}
{\Large\bf  Black Holes as Critical Point of  Quantum Phase Transition} 

\end{center}

\vspace{0.1cm}

\begin{center}

{\bf Gia Dvali}$^{a,b,c,d}$\footnote{georgi.dvali@cern.ch} and {\bf Cesar Gomez}$^{a,e}$\footnote{cesar.gomez@uam.es}

\vspace{.6truecm}


{\em $^a$Arnold Sommerfeld Center for Theoretical Physics\\
Department f\"ur Physik, Ludwig-Maximilians-Universit\"at M\"unchen\\
Theresienstr.~37, 80333 M\"unchen, Germany}


{\em $^b$Max-Planck-Institut f\"ur Physik\\
F\"ohringer Ring 6, 80805 M\"unchen, Germany}

{\em $^c$CERN,
Theory Department\\
1211 Geneva 23, Switzerland}


{\em $^d$Center for Cosmology and Particle Physics\\
Department of Physics, New York University\\
4 Washington Place, New York, NY 10003, USA}

{\em $^e$
Instituto de F\'{\i}sica Te\'orica UAM-CSIC, C-XVI \\
Universidad Aut\'onoma de Madrid,
Cantoblanco, 28049 Madrid, Spain}\\

\end{center}


\begin{abstract}
\noindent  
 
{\small
 We reformulate the quantum black hole portrait in the language of modern condensed matter physics. We show that black holes can be understood 
 as a graviton Bose-Einstein condensate at the critical point of a quantum phase transition, identical to what has been observed in systems of cold atoms. 
  The Bogoliubov modes that become degenerate and nearly gapless at this point are the holographic quantum degrees of freedom responsible for 
the black hole entropy and the information storage.  They have no 
(semi)classical counterparts and become inaccessible in this limit.  These findings indicate a deep connection between the seemingly remote systems and suggest a new quantum foundation of holography.  They also open an intriguing possibility of simulating black hole information processing in table-top labs.  }

\end{abstract}

\thispagestyle{empty}
\clearpage

\section{Introduction}

Exchange of ideas between condensed matter and particle physics 
has a long history. 
In the present paper we would like to establish one more link: A fundamental  connection between the 
physics of black holes and critical phenomena in quantum Bose-Einstein condensates (BEC) that are  formed in ordinary quantum systems such as cold atoms and photon gasses.   

  This connection originates from a recently formulated quantum theory of black holes, according to which black holes represent quantum  Bose-Einstein condensates of gravitons \cite{Nportrait}. 
  
   In the usual treatment gravitational systems, such as the black holes (or even an entire Universe),  are introduced through the background geometry that they 
   produce.  Thus, in the semi-classical approximation, one studies small fluctuations 
about the background, but the background geometry itself is treated as an intrinsically 
classical entity.  However, in nature there are no true classical objects, the Planck's constant is non-zero, $\hbar \, \neq \, 0$.  So in the semi-classical treatment we are working in the limit in which the quantum constituents of the geometric background cannot be resolved. 

  Hence, what are the true quantum constituents of the classical geometry? 
  
 Just in the same way as a laser beam is an emergent classical description 
 of a large occupation number of photons,  the classical geometry must be handled as an effective description of a quantum state with large graviton
 occupation number.  When such state is a ground-state, the gravitational field 
 is effectively a Bose-Einstein condensate. 
 This is the case for black holes \cite{Nportrait}.  
  
 Unlike the photons (which are electrically neutral) gravitons gravitate and can form a self-sustained 
 Bose-condensate,  a  black hole.  The special property of such a condensates is that they are at the point of {\it maximal packing}.  The maximal packing  
 means that the size of the condensate, $L$, depends on the occupation number $N$ in such a way that it is impossible to further increase $N$ without increasing 
$L$.   Essentially, at the point of maximal packing $N$ becomes the sole characteristic of the system.   In particular, the size scales as
$ L\, = \, \sqrt{N} \, L_P$, whereas the coupling between individual particles scales as $\alpha \, = \, 1/N$.  Putting it simply, a black hole represents a large-$N$ system in the 't Hooft's sense  \cite{tHooft}, with the critical value of the coupling, $\alpha N \, = \, 1$.

 This picture naturally explains all the semi-classical   
mysteries of black holes. Particularly, the Hawking radiation\cite{hawking} and the negative specific heat result from quantum depletion of the condensate. 
The spectrum of radiation is thermal up to $1/N$-corrections with effective temperature $T\, = \, \hbar /\sqrt{N} L_P$.  The resulting heat capacity is obviously negative, since $N$ decreases as a result of the depletion. 

It is interesting, that emergence of thermality has nothing to do with the temperature of the condensate, but instead results from the self-similarity of the depletion and leakage, which does not change the $N$-dependence of the black hole characteristics.  

 Another black hole mystery is the origin of Bekenstein entropy \cite{Bekenstein}
 and the quantum mechanism of information-storage and processing by a black hole.  The scaling of Bekenstein entropy as  the horizon area  
 $S \, \sim \, L^2/L_P^2$, creates the impression that the horizon is an union of 
 $N$ Planck-size pixels each housing a {\it distinguishable} degree of freedom that can be in a discrete number of degenerate states (e.g., $\pm$) resulting into an exponentially  large number of micro-states.  Due to this property,  these hypothetical degrees of freedom are often called {\it Holographic } \cite{hologram}. 
 
 In the semi-classical picture the microscopic origin of these degrees of freedom 
 is mysterious (and as we shall show, inaccessible in principle).  Instead in our quantum picture, these degrees of freedom naturally originate as collective quantum excitations of the graviton Bose-condensate, 
huddled within a mass gap of order $1/N$.  In \cite{Nportrait} we refer to them as {\it flavors}.  These flavor degrees of freedom are intrinsically-quantum 
and they decouple as $1/N$ in the classical limit, as they should. Correspondingly 
in the (semi)classical limit it becomes infinitely hard to resolve them.  
This explains why in this limit black holes can store an arbitrary amount 
of information, without ever releasing it. 

 If black holes are Bose-Einstein condensates then it is natural to expect that at least some 
 of its properties must have counterparts in ordinary Bose-Einstein condensates\cite{review} such 
 as in systems of cold atoms \cite{ueda, ueda1}  or photons \cite{photon}. 
 The purpose of this paper is to establish this connection. 
The motivation for such an analysis is pretty clear. 
First, it is of fundamental importance to establish unity of physical phenomena in seemingly remote systems.  Secondly, such a connection can potentially 
enable us to simulate quantum black hole physics in the table-top labs. 
 A potential byproduct of such a simulation can be the use of black hole information storage and processing in cold atomic or photonic systems.    
  
   In this paper we shall try to make a first step in this direction and to show  
 that the connection is much deeper that what one would have naively thought.   

  To summarize our findings shortly: 
  
  $~~~~$ 
  
   { \it Black holes represent  Bose-Einstein-Condensates of gravitons at the critical point of 
   a quantum phase transition.} 
    
   $~~~$ 
   
    Quantum phase transitions are well-known in condensed matter physics (see e.g., \cite{schadev}).   The category of quantum  transitions 
  that is our main focus was studied recently in cold atomic systems \cite{ueda, ueda1}.  As we shall explain, such phase transitions capture the key physics of the black hole quantum portrait. 
  
   The essence of the connecting phenomenon 
is the following.  In the presence of an attractive interaction a fixed size BEC undergoes a phase transition 
 above critical values of the occupation number $N$.  The uniform BEC becomes 
 unstable and moves into a phase of a bright soliton.  
  At the critical point Bogoliubov modes become almost degenerate with the ground-state, within the energy gap that collapses as $1/N$. These gapless modes reflect the underlying breaking of symmetry and the corresponding appearance of Goldstone modes.  At the same point 
  the quantum depletion of the condensate becomes important. 
  
     Detailed comparison of the above system with the black hole picture of \cite{Nportrait} reveals that 
     all the above phenomena have exact counterparts there. 
   Namely, the critical value of the occupation number in the black hole case 
   corresponds to the point of maximal packing (the critical value of the 't Hooft's coupling) or equivalently to self-sustainability 
   of the graviton condensate.  The gapless Bogoliubov  modes are playing the role 
   of the holographic flavors that account for the black hole entropy and the quantum depletion is the Hawking radiation.    
  The brief summary of the black hole - BEC correspondence is: 
  
  \begin{itemize}
  \item Maximal packing (self-sustainability)  $\, \leftarrow\rightarrow \, $ Critical point of a quantum phase transition  
  \item Holographic degrees of freedom (flavors) $\, \leftarrow\rightarrow \, $ Gapless Bogoliubov modes at the critical point
  \item Bekenstein entropy $\, \leftarrow\rightarrow \, $ Quantum degeneracy of the BEC state at the critical point
  \item Hawking Radiation $\, \leftarrow\rightarrow \, $ Quantum depletion and leakage of the BEC 
\end{itemize}

 It is important to stress, that we are not dealing with a crude analogy, but with 
 identical physics. 
 Of course, black holes have some peculiar characteristics (so far) not exhibited 
 by ordinary lab condensates.   In the lab systems the critical point is achieved by tuning the external factors (e.g., size of the system, number of 
 atoms and interaction strength). Due to this the depletion puts them away 
 from the critical point. 
 
 In contrast, black holes are self-tuned and always stay at the critical point 
 due to the self-similarity of the depletion.  Emission of a quantum takes a black hole of graviton number $N$ into the one with $N-1$ with all the characteristics 
depending on $N$ self-similarly.   Analogous effect could in principle  be achieved in the lab if one externally adjust parameters in order to track the depletion.  

  As a final remark,  we have suggested that the connection between the maximal packing and the holography must not be limited to the black hole case and must be more general.  In particular, as a supporting evidence, by  extending this connection to an AdS geometry and treating it as a graviton condensate we have observed \cite{Nportrait} that it also appears at the point of maximal packing. Moreover, the occupation number $N$
  of gravitons in AdS exactly reproduces the central charge of the CFT that is independently predicted by AdS/CFT correspondence \cite{maldacena}.  Is this a simple coincidence? 
  
   The results of the present paper suggest that it is not.  It originates from the 
general feature of an overpacked BEC being at the critical point, where 
Bogoliubov modes become degenerate and the system 
effectively becomes conformal.    We therefore suggest: 

$~~~~$

{\it The understanding of classical geometries as BECs at the critical point provides  a quantum foundation of holography.}

 $~~~~~$ 
 
   The maximally packed systems ($\alpha N = 1$), such as black holes or  AdS space, are equivalent to BEC's at the critical point of a quantum phase transition, and as such they are described by (nearly conformal) physics 
  of degenerate Bogoliubov modes.  The degree of conformality should 
  be determined by the depletion properties of the condensate.

  The paper is organized as follows. 

  In the next section we briefly review the essentials of the black hole quantum 
  portrait in terms of  BEC of gravitons.   In section 3 we make contact
  between this picture and the quantum phase transition in BECs appearing in condensed matter and atomic  systems\cite{ueda, ueda1} and show that they are governed by the same physics. 
  In order to establish this connection we study a prototype theory of BEC that exhibits critical transition and show that these properties closely match the ones of graviton BEC in the quantum picture of a black hole\cite{Nportrait}.    In section 4 
  we discuss the peculiarity of the graviton BEC, that allows it to be stuck 
  at the critical point, even during the quantum collapse and leakage.   
 In section 5 we discuss why maximal packing (equivalently critical point) 
 of BEC is crucial for allowing large entropy and information storage.   
  In section 6 we discuss the generalization of our results to other systems with maximal packing and suggest that this provides an universal quantum foundation of holography.   Finally, we discuss some potential implications of our results and give an outlook. 
  We shall set the speed of light to one, but keep $\hbar$ explicit.  
  We shall ignore all the irrelevant numerical factors. 
  
  \section{Black Holes as Bose-Einstein Condensates}
   
   In this section we shall briefly discuss some essentials of the black hole 
   quantum portrait in order to prepare a ground for establishing the connection with
   condensed matter systems.   For a more detailed discussion the reader is referred to the original papers \cite{Nportrait}.  
 
   Gravitons are massless spin-2 particles. The strength 
  of graviton-graviton interaction is measured by a dimensionless coupling "constant", 
  \begin{equation}
   \alpha\, \equiv  \,  {L_P^2 \over L^2} \, , 
   \label{alpha} 
   \end{equation}
where, $L$ is a characteristic wave-length of the gravitons participating in the 
interaction and $L_P$ is the Planck length.
In terms of Newton's gravitational constant, $G_N$, the  Planck length 
is defined as $L_P^2  \, \equiv  \,  \hbar \,  G_N$.   The physical meaning of the 
above coupling can be understood in  simple terms as the relativistic generalization 
of the Newtonian attraction among two gravitons. Notice that the latter force 
acting among two non-relativistic massive particles of mass $m$ can be written in terms of $\alpha$ as, 
   \begin{equation}
   V(r)_{Newton}  \, = \, - \hbar {\alpha\, \over r} \, , 
   \label{alphamassive} 
   \end{equation} 
 with the same $\alpha$ given by (\ref{alpha}), but with the only difference that for a massive particle 
 $L$ has to be understood as its Compton wave-length, $L \, \equiv \, {\hbar \over m}$. The difference for gravitons is that, because they are massless, the role of the Compton wave-length is replaced by an actual wave-length.  
 
   From equation (\ref{alpha}) it is obvious that if wave-lengths are large, 
   the interaction among the gravitons is extremely weak.   For example, 
   for gravitons of wavelength $L \, = \, 1$cm, the quantum interaction strength 
  is $\alpha\, = \, 10^{-66}$!   One would say that for all practical purposes 
  such gravitons should behave as free. However, gravitons are bosons, and because of this their occupations numbers can be very high. 
 In such a case the collective effects become extremely important. 
   The key point of our theory is that gravitons can {\it self-condense} and this 
 condensates are black holes.   As we shall see, because of the nature of the coupling (\ref{alpha}) this condensates are 
 very special as they are always at the critical point.   
  
   In order to see this let us imagine that we wish to localize as many soft gravitons as possible within a space region of size $L$.  In other words we 
   are trying to form a BEC of gravitons of characteristic wave-length $L$ by gradually increasing the occupation number $N$. 
   At the beginning, when  $N$ is small, graviton interactions are negligible and 
 we need external sources to maintain the condensate. 
 So for small occupation numbers, the behavior is similar to a photon condensate, which requires external binding potentials. 
 However, as we increase $N$ the effects of the interaction 
 become dramatic.  Individual gravitons  feel stronger and stronger 
 collective binding potential and for the critical occupation number, 
 \begin{equation}
 N \, = \, N_c \, =\, {1 \over \alpha} \, ,
 \label{cr}
 \end{equation}
 the graviton condensate becomes {\it self-sustained}. 
 This self-sustainability condition can be obtained by equating the kinetic energies of individual gravitons, $E_k \, = \, \hbar /L$, with the collective binding potential, $V\, = \, - \alpha \, N \, {\hbar \over L}$,
 \begin{equation}
 E_k \, + \, V \, = \, \left (1 \, - \alpha\,  N \right ) \, {\hbar \over L} \, = \, 0 \, ,
 \label{self}
 \end{equation}
 which is satisfied for the critical value of $N$ given by  (\ref{cr}). 
 
  An extremely important property of  the critical point is that it also corresponds 
  to the point of {\it maximal packing}.   The concept of maximal packing is that the system 
  is so  densely packed that its defining characteristics becomes $N$. 
 In particular, 
 \begin{equation}
  L \, = \, \sqrt{N} L_P \, , ~~~ \, \alpha \, = \, 1/N \, . 
  \label{overpack}
  \end{equation}
 But, for gravitons being at the overpacked point also means that further  increase of $N$ without increasing $L$ becomes impossible.  Any further increase of $N$ results into the  increase of the wave-length in such a way 
 that the system stays at the maximal-packing point  (\ref{overpack}). 
 
   Notice, that equation (\ref{self}) clearly indicates that the critical point (\ref{overpack})  can be achieved for arbitrary $N$, but it is not enough to see
 why $L$ cannot decrease beyond $L\, < \, \sqrt{N}L_P$. 
 Naively,  such a decrease of $L$ would results into an even stronger bounded system. 
 
  Such collapse of $L$ indeed takes place, but remarkably it cannot take the system out of the critical point (\ref{overpack}).   The reason is that the decrease 
  of $L$ is balanced by the decrease of $N$ due to quantum depletion 
 and leakage of the condensate. As a result the condensate 
 slowly collapses, but it looses gravitons at the same rate. In this way,  the system always stays at the critical point (\ref{overpack}).

   The reason for the diminishing of $N$ is that the graviton condensate undergoes a quantum depletion and the depleted quanta leak out.  
 The key of this phenomenon is that due to the interaction with their 
 fellow gravitons, some of the bosons get excited above the ground-state. 
 But, since the ground-state energy is within $1/N$ from the free-escape point, 
 the excited gravitons that gain energies above this tiny gap leave the condensate and join continuum.  In other words, the condensate starts to leak, with a depletion rate which is essentially given by   
\begin{equation}
 \Gamma_{leakage} \, = \,   {1 \over  \sqrt{N} L_P}   \, + \,  L_P^{-1} \,  {\mathcal O} (N^{-3/2}) \, . 
\label{leakage}
\end{equation}
 This rate can be easily understood from the following estimate. 
 Since the graviton-graviton coupling in the condensate 
 is $1/N$, the probability for any pair of gravitons to scatter  is suppressed by $1/N^2$, however this suppression  is compensated by a combinatoric factor  $\sim \, N^2$  counting the number of available graviton  pairs.  As a result,  the rate of the graviton emission from the condensate is simply given by 
 the characteristic energy of the process (inverse wave-length of gravitons).

     The above quantum depletion translates into the following leakage law,     
     \begin{equation}
   {\dot N} \, = -  {1 \over \sqrt{N} L_P}  \, + \, L_P^{-1} \,  {\mathcal O} (N^{-3/2}) \, , 
   \label{deplete}
   \end{equation}
 where dot stands for the time-derivative. 
   
  It is precisely this quantum leakage of the graviton BEC that  (only!) in 
the semi-classical limit becomes Hawking radiation.  
  
 The correct understanding of the semi-classical limit is the key for  
 understanding  why all the above quantum physics of graviton BEC was missed 
in the previous analysis.   
  
  The semi-classical limit is defined by the following double scaling limit, 
  \begin{equation}  
   N \, \rightarrow \, \infty, ~~~L_P\, \rightarrow \, 0\,,   ~~~ L\equiv \sqrt{N}L_P \, =\, {\rm finite}, ~~~\hbar \, = \, {\rm finite}\, . 
   \label{thelimit}
   \end{equation}

 Thus, in the language of BEC the semi-classical limit is the limit in which 
 all the quantum physics of the condensate decouples as $1/N \, \rightarrow \, 0$ and becomes impossible to resolve.  What was a quantum condensate 
 now becomes a collection of infinite number of infinitely soft  non-interacting 
 bosons all the individual identities of which are lost. 
  All the semi-classical black hole mysteries are a direct consequence of this 
 unphysical  treatment.  One of the consequences is the exact thermality of 
Hawking radiation.

   This immediately follows from the  leakage law. Which in this limit 
(by rewriting $N$ in terms of the black hole mass) becomes the 
Stefan-Boltzmann law for a black hole with  Hawking temperature given by  $T \, =  \, {\hbar  \over L}$, 
    \begin{equation}
   {\dot M} \, = -  {\hbar  \over  L^2}  \, . 
   \label{depletemass}
   \end{equation}

Notice, that the exponential suppression  of higher frequencies, usually attributed to the thermality  of the source,  follows from the combinatorics of the quantum depletion.  For example, emission of a graviton with much shorter 
wavelength,  $ \sim  k^{-1} \sqrt{N}L_P$ (where $k \, \gg \, 1$ is a parameter ), 
requires a re-scattering cascade process of at least $k$ gravitons in the condensate. 
Due to the variation of the effective graviton coupling along the cascade the corresponding rate for large $k$ is suppressed by the factor, 
\begin{equation}
 \Gamma_{k >> 1} \, \propto \,  N^{-k} \, k! \, ,  
 \label{higher}
 \end{equation}
 where the extra $k!$ comes from the correction to the graviton coupling for a cascade taking place in $k$ consecutive steps. When $k$ is a fraction of $N$, 
 the suppression factor goes like $e^{-k(1\, + \, ln N/k)}$. 
In the semi-classical limit (\ref{thelimit}) the above suppression  reproduces 
 the exponential Boltzmann-type suppression,  which is typical of the thermal spectrum. 
 Nonetheless the underlying quantum physics of this thermal-like spectrum has nothing to do with the thermality of the source, since condensate 
 is in fact cold, but with the underlying quantum physics of BEC being at the 
 overpacked critical point. We will elaborate further this discussion in the next section after introducing a concrete microscopic model of the Bose condensate.

  \section{Black Hole as BEC at Quantum Phase Transition}   

 We now wish to establish the connection between the above-discussed black hole quantum portrait and the critical phenomena in ordinary BEC, such as were 
 observed in cold atoms in \cite{ueda, ueda1}.  However, since we would like 
 to keep our discussion as general as possible, we shall consider a simple prototype model that captures the key features of the phenomenon.
   Let $\Psi(x)$ be a field operator describing the order parameter 
 of a Bose-gas.  The particle number density is given by the correlator
 $n(x) \, = \, \langle \Psi(x) \Psi(x) \rangle$.   The simplest hamiltonian that 
 takes into the account the self-interaction of the order parameter  can be written in the form, 
 \begin{equation}
  H\, = \, -\, \hbar L_0  \int \, d^3x \Psi(x) {\bf \nabla}^2 \Psi(x) \, - \, g \,   
  \int \, d^3 x \, \Psi(x)^+\Psi(x)^+\Psi(x)\Psi(x) \, , 
  \label{hamilton}
  \end{equation}
  where, $L_0$ is a parameter of length-dimensionality, and $g$ is an interaction coupling constant of dimensionality [length]$^3\times$[mass].   Since we are looking for a connection with gravity we assume the interaction to be attractive. 
  We shall put the system in a finite box of size $R$ with periodic boundary conditions $\Psi(0) \, = \, \Psi(2\pi R)$ and with the total particle number being  $N$. This implies the normalization condition, 
\begin{equation}
 \int \, d^3 x \,  \Psi^+ \Psi\, = \, N \, .
 \label{norm}
 \end{equation} 
 Performing a plane-wave expansion,  $\Psi \, = \, \sum_{\bf k} \, {a_{\bf k} \over \sqrt{V}} 
 e^{i {{\bf kx} \over R}}$ where $V=(2\pi R)^3$ is the volume and 
 $a_{\bf k}, a_{\bf k}^+$ are particle creation an annihilation operators, 
 $[a_{\bf k} a_{\bf k'}^+] \, = \, \delta_{{\bf k k'}}$, we can rewrite the Hamiltonian 
 as    
  \begin{equation}
  \mathcal{H} \, = \, \sum_{{\bf k}} \, {\bf k}^2 a_{\bf k}^+a_{\bf k}
  \, - \, {1 \over 4} \, \alpha \,   \sum_{\bf k} \, a_{\bf k + p}^+  \, a_{\bf k' -p}^+ \, a_{\bf k} \, a_{\bf k'} \, , 
  \label{hamiltonk}
  \end{equation} 
   where $\alpha \, \equiv \,  {4g R^2  \over \hbar V L_0}$ and  
  $\mathcal{H}\, \equiv \, {R^2 \over \hbar L_0} \, H$.

   We shall now study the spectrum of low lying excitations about an uniform BEC.  That is, we 
   assume that most of the particles occupy the $k =0$ level 
   and study the small quantum fluctuations about this state. The spectrum of fluctuations is determined by the Bogoliubov-De Gennes equation. In first approximation we can use the 
  Bogoliubov replacement 
  \begin{equation}
   a_0^+ \, = \, a_0 \, = \, \sqrt{N_0} \, \simeq \, \sqrt{N} \, , 
   \label{bogoliub}
  \end{equation}
  of the ground state creation annihilation operators into classical c-numbers. Note that this approximation relies on taking $N\, \gg \, 1$ while keeping $\hbar$ different from zero.
 Keeping only terms up to quadratic order in $a_{\bf k}^+, a_{\bf k}$ for  ${\bf k}  \, \neq \, 0$, and taking into account the normalization condition (\ref{norm}), 
  \begin{equation}
  a_0a_0 \, + \, \sum_{\bf k \neq 0} \, a_{\bf k}^+ a_{\bf k} \, = \, N \, .
  \label{norma}
  \end{equation}
  leads (up to a constant ) to the following Hamiltonian describing the small quantum fluctuations 
  
    \begin{equation}
  \mathcal{H} \, = \, \sum_{{\bf k \neq 0 }} \left ( {\bf k}^2 \, + \,  \alpha N \right/2 ) 
  \,  a_{\bf k}^+a_{\bf k}
  \, - \, {1 \over 4} \,  \alpha \, N \sum_{\bf k \neq 0} \left ( a_{\bf k}^+  \, a_{\bf -k}^+ \, + \,  a_{\bf k} \, a_{\bf -k} \, \right ) \, . 
  \label{hamiltonk}
  \end{equation}
  In order to diagonalize the hamiltonian we perform a Bogoliubov transformation, 
  \begin{equation}
  a_{\bf k} \, = \, u_{\bf k} \, b_{\bf k} \, + \, v_{\bf k}^* b^+_{\bf k}  \, .
  \label{btrasform}
  \end{equation}
  The  Bogoliubov coefficients are given by , 
  \begin{equation}
  u,v \, = \, \pm  {1 \over 2} \left ( {{\bf k}^2 \, - \, \alpha N/2 \over \epsilon ({\bf k}) } 
  \pm \, 1 
  \label{uv} \right ) \, ,
  \end{equation} 
 leading to the following spectrum of the Bogoliubov modes, 
 \begin{equation}
 \epsilon({\bf k}) \, = \, \sqrt{ {\bf k}^2 ( {\bf k}^2 \, - \, \alpha N)} \, .
 \label{bspectrum}
 \end{equation}
 The Hamiltonian in terms of $b$-particles is diagonal and has the following form
  \begin{equation}
  \mathcal{H} \, = \, \sum_{{\bf k}} \epsilon ({\bf k}) \, b_{\bf k}^+ b_{\bf k} \, + \, 
  {\rm constant} \, .
  \label{bhamilton}
\end{equation} 
   As it is clear from (\ref{bspectrum}) the first Bogoliubov energy vanishes 
 for 
 \begin{equation}
 N \, = \,  N_c  = \,  {1\over \alpha} \, 
 \label{ncrit}
 \end{equation}
 and the system undergoes a quantum phase transition.
  This is exactly the phase transition observed in \cite{ueda, ueda1}. 
   The essence of this phase transition is that for  $N \, > \, N_c$  the first Bogoliubov level becomes tachyonic and the uniform  BEC is 
 no longer a ground-state.   
     Taking into the account ${1 \over N}$-corrections, it is clear that 
 the gap between the uniform ground-state and the Bogoliubov modes collapses to $1/N$ and becomes extremely cheap to excite these  modes. So by quantum fluctuations the system starts to be populated by Bogoliubov modes 
 very easily.   This means that the condensate starts to undergo a very efficient 
 quantum depletion.  The number density of the depleted $a$-particles 
 to each ${\bf k}$-levels are given by  
 \begin{equation} 
   n_{\bf k} \, = \, |v_{\bf k}|^2 \,.   
 \label{av}
 \end{equation} 
   Since $n_{\bf k}$ decreases as $1/|{\bf k}|^4$ for large $|{\bf k}|$, the 
 total number of depleted particles is well-approximated by the first-level 
 depletion, which gives, 
 \begin{equation}
 \Delta N \, \sim \, n_1 \, = \, \left ( { 1 \,  - \,  \alpha \, N /2  \over \sqrt{ 1 \, - \, 
  \alpha \, N}} \, - \, 1 \right ) \, \simeq \, \sqrt{N}  \, .
  \label{n1}
  \end{equation}
   The striking similarity of the above BEC physics with the black hole quantum portrait 
   suggest that in both cases  we are dealing with one and the same physics of the quantum phase transition.  Indeed the physics of the graviton condensate is reproduced for the particular case of $L_0 \, = \, R \, = L$ and  $g = \hbar \, L_P^2$.

   The criticality condition (\ref{ncrit}) then is nothing but the self-sustainability 
 condition (\ref{cr}) which implies that the graviton condensate is maximally packed
 (\ref{overpack}).  
   The energy gap to the first Bogoliubov level is then given by 
 \begin{equation}
 \epsilon_1 \, = \, { \hbar \over L \sqrt{N}} \, = \, {1 \over N} {\hbar \over L_P} \, .  
\label{gap}
\end{equation}
 This  expression summarizes a remarkable property of maximally-packed 
 systems: 
 
  {\it The energy cost of a collective excitation can be made arbitrarily low 
 by increasing the occupation number of bosons in the BEC.}    
 
 Thus, by increasing $N$ one can encode an essentially-unlimited amount of information in these modes.  Notice, that in semiclassical limit 
 (\ref{thelimit}) the energy gap collapses to zero and BEC (black hole) becomes 
 an infinite capacitor of information storage!   

  What we are uncovering is that this is a very general property 
 of overpacked BEC's  which are at the critical point of a quantum phase transition.  
  In both cases, the cold atomic system of \cite{ueda, ueda1} versus the graviton condensate,  the  key point is the maximal packing.
  The  overpacking of the system results into the collapse of the mass gap 
 and the Bogoliubov modes become degenerate within a  $1/N$-window. 
   These almost-degenerate Bogoliubov modes are the quantum  holographic degrees of freedom (flavors) that are responsible both for the entropy as well as for the efficient depletion of the system. Notice, that these 
 degenerate Bogoliubov modes are intrinsically-quantum and have no classical counterparts. In the classical limit they decouple as $1/N$ and become unobservable.  
 
 The way the BEC state acquires entropy at the critical point is easy to figure out. In fact in the homogeneous phase for $N<N_c$ the low lying states in the Fock space are characterized by $|n,N-n>$ where $n$ represents the number of quanta in the first excited state and $N-n$ the ones in the condensate ground state. The quantum phase transition takes place when $N=N_c$. The specific feature of the quantum phase transition is that all these excited states in the Fock space become quasi degenerate ( at $\frac{1}{ N}$ order ) in energy manifesting the underlying spontaneous breaking of symmetry and the appareance of a Goldstone mode. Since at the critical point the number of quasi degenerate ground states is $O(N)$ we can effectively define $\sim N$ Bogoliubov quasi zero modes. 
 In the presence of any additional discrete characteristics of bosons (e.g., such as helicity) the scaling of entropy as $S \, \sim \, N$ is a natural expectation.  
 
 In this qualitative approach we do not attempt to get the numerical coefficients. Our target instead is to uncover the quantum physics behind the black hole entropy as the quasi degenerate nature of the corresponding BEC state at the quantum critical point. In terms of information theory what we observe is that once we reach the quantum critical point we can use the Bogoliubov quasi zero modes to store information at a minimal cost of energy.
\footnote{In the previous paragraph we have simply defined the black hole entropy as measuring the degeneration of the corresponding BEC ground state at the critical point. In addition to this microcanonical notion of entropy we can naturally define a quantum contribution. Denoting $|BEC(i)>$ with $i=1,..N$ the quasi degenarate ground states at the critical point we should think of the black hole quantum state as some quantum superposition $|BH>= \sum_ic_i|BEC(i)>$ with $\sum_i|c_i|^2=1$ and the corresponding density matrix as $\rho =|BH><BH|$. If we wash out the off diagonal pieces of $\rho$ as we do when we measure the system we gain an amount of Von Neumann entropy of the order $-\sum_i (c_i^2lnc_i^2) \sim  lnN$. This is the quantum contribution to the black hole entropy in the BEC portrait.} 
 
  Let us now derive the black hole leakage law (\ref{deplete}) from the Bogoliubov treatment of the BEC at the critical point.  The equation (\ref{n1}) 
 gives the number of the depleted particles in the absence of back reaction. 
  Since we are interested in time dependence of $N$, we need to divide 
 the number of depleted particles $\Delta N \sim  \sqrt{N}$ by minimal time $\Delta t$ that such depletion takes. This time is given by the time that it takes 
 $\Delta N$ number of particles to re-scatter.   The time for re-scattering of a single pair is given by 
 \begin{equation}
 \tau \, \sim \, L \, \alpha^2 \, N^2 \, \sim \, \sqrt{N}L_P \, .
 \label{singlet}
 \end{equation}   
 Correspondingly,  the 
 time for $\sqrt{N}$ such re-scatterings is $\Delta t \, =  \, \sqrt{N} \tau \, = \, N L_P$. Thus,  the resulting leakage law up to $1/N$ corrections is, 
      \begin{equation}
   {\dot N} \, =  \, -\,  {\Delta N \over \Delta t}  =  - \, {\sqrt{N} \over N L_P} \, =    
   -   { 1 \over \sqrt{N} L_P} \, , 
   \label{deplete1}
   \end{equation}
 which exactly reproduces (\ref{deplete}).   Thus, we have reproduced the black hole evaporation law from the depletion of the cold Bose-Einstein condensate at criticality. At this point it is interesting to observe that the re-scattering time defined above coincides with the causal cell for a speed of {\it sound} $\alpha^2 N^2$ equal to the speed of light. This is again a typical property of the quantum critical point that very likely lies at the origin of fast scrambling \cite{scrambling}. 
  
   Notice, that the value we have used for $\Delta N$ was derived in the absence of back reaction. The peculiarity of the graviton condensate allows us to neglect this back reaction.  The reason is that black hole graviton condensate is always 
at the critical point since $\alpha$ is tracking $1/N$.   So approximation of no-back  reaction is always good up to $1/N$.  This is why $\dot{N}$ is very well approximated by $\Delta N / \Delta t$. 
  The situation is different for the cold atomic systems, where $\alpha$ is an external parameter and one has to take into the account the back reaction, as it was done in \cite{ueda}.  
   If $\alpha$ is not tracking $N$, then the change of $N$ by $\Delta N$ offsets $\alpha N$ by  $\alpha \Delta N$ and in equation (\ref{n1}) one has to replace 
   $\alpha N \, \rightarrow \,  \alpha N ( 1 \, + \, \Delta N/N)$, which gives       
   $\Delta N \, \sim  \, N^{1/3}$.  Obviously the black hole quantum phase transition should be characterized by some {\it critical exponents} roughly characterizing the holographic CFT. However the Bogoliubov approximation we are considering here is simply equivalent to a mean field approximation.
   
   Before closing this section it would be illustrative to compare the BEC derivation of the depletion with the semiclassical derivation of Hawking radiation in the black hole geometry. In both cases the essence of the derivation lies on the Bogoliubov transformation. In the black hole case and simplifying things representing the near horizon as Rindler geometry, the relevant transformation connects creation annihilation operators relative to Minkowski and Rindler vacua. This transformation leads to a typical thermal spectrum $\Delta(N_{\omega}) = (e^{\omega T}-1)^{-1}$ with $T$ the Hawking temperature. In the IR we get $\Delta(N)\sim \frac{T}{\omega}$, while in the UV we get the typical thermal exponential suppression. In the case of the BEC we have derived above the depletion in the IR regime obtaining $\Delta(N) \sim \sqrt{N}$. Nicely enough this corresponds to the minimal energy $\omega \sim \frac{1}{N}$ in the BEC and to an effective temperature $T \sim \frac{1}{\sqrt{N}}$ in agreement with the depletion law. Moreover the exponential suppression in the UV can be easily understood once we have uncover the meaning of the black hole entropy. In fact when we consider the emission of a very energetic quanta, we are forced to build up these quanta with a certain number $k$ of soft quanta occupying the 
 ground-level of the condensate.   
  This effectively reduces the degeneracy of the BEC ground state by a factor of order $e^k$. In other words when the system emits a hard quantum the price to pay is to reduce accordingly the multiplicity of quasi-degenerate Bogoliubov modes.

   \section{Being Stuck at the Critical Point} 
   
   As we said above, the important property of the graviton BEC is the impossibility 
   to enter into the strong coupling regime, $\alpha N \, \gg \, 1$.
  Despite the fact that  black holes deplete and leak gravitons they always remain at the critical point, because leakage is self-similar in $N$.   
     
   In order to understand this peculiarity, let us first discuss what is happening in 
   other systems for which entering into the  strong coupling regime is possible.    
  For cold atomic BEC's discussed in \cite{ueda},  the quantum phase transition signals the formation of  a {\it bright soliton}.   That is,  an overpacked condensate ($N \, \gg \, N_c$) prefers to store particles non-uniformly and store them in higher momentum modes.  The critical point (\ref{cr}) marks the transition between the two regimes. The reason we can ,in this case, enter into the strong coupling regime is because although the attractive interaction increases we can compensate it by the quantum presure created by  higher momentum modes in the band of states that are quasi degenerate in energy with the uniform BEC ground state.  
  The soliton configuration that represents a ground-state in this regime and can be well-approximated by a  localized solution of the Gross-Pitaevskii equation,  
  \begin{equation}
  i \hbar {\partial \Psi \over \partial t} \, = \, - \,  \left ( \hbar L_0 \, { \bf \nabla}^2  \Psi \, +  \, 2 g  \, \Psi^+ \, \Psi 
  \right ) \Psi \, . 
  \label{GP}
  \end{equation}. 
  
   In one space dimension for $\alpha N \, \gg \, 1$ this equation has a well-known exact solution \cite{soliton},  a bright soliton,  
 \begin{equation}
 \Psi_s(x) \, \propto  \, \sqrt{\mu \over g} \,  {\rm sech} \left ( \sqrt{{\mu  \over \hbar L_0}}
 (x - x_0) \right )  \, ,     
   \label{soliton} 
   \end{equation}
  where $\mu$ is a chemical potential, which from normalization condition 
  scales as $\mu \, \sim \, g^2N^2/\hbar L_0$. Therefore, the argument of 
  the (\ref{soliton}) scales $\sim (\alpha N) x/R$.
   
   This system exhibits a Goldstone zero-mode (corresponding to a spontaneous breaking of translational invariance) and higher excitations (breathing modes ) separated by an energy gap. 
   
   In  three dimensions however the solitons are unstable and collapse.
  This can be understood from the following simple energetics argument. 
 Consider a deformation of the uniform condensate such that 
 we localize most of the particles  within a region of size $L$. 
 That is, the order parameter $\Psi$ is localized within the region $L$.  
 Due to the normalization condition (\ref{norm}) the over-density scales as $|\psi |^2  \sim N/L^3$. 
  The energy corresponding to such a configuration from Hamiltonian (\ref{hamilton}) is,  
\begin{equation}
 E \, \sim  \,  {\hbar L_0} {N \over L^2} \,   - \,  g {N^2 \over L^3} \, .   
 \label{deltae}
 \end{equation}
 This $L$-dependence has no minimum.  It has an extremum at the critical point 
 $\alpha N$ = 1.  
 Beyond the critical point $\alpha N \, > \, 1$ the system collapses towards $L \, \rightarrow \, 0$.  The collapse indicates that the system prefers to store 
 more and more quanta into the higher momentum modes and the condensate
  is getting more and more localized.   
 
  What is the connection of this phenomenon to our picture of a black hole?  
 The peculiarity of the black hole graviton condensate is that although it also collapses the collapse takes place through a cascade of successive condensates $N\rightarrow N-1 \rightarrow N-2 ..$,  all of them at the critical point ! This is due to the fact that the 
 black hole collapses  by a quantum depletion which is accompanied by the  leakage.  Depleted quanta escape and leave the condensate, decreasing 
 $N$ according to (\ref{deplete}).  
  
  In other words, the quantum-mechanical collapse of the graviton condensate 
  is simply a quantum progenitor of what in semi-classical limit becomes the Hawking radiation \footnote{It is customary to try to describe black hole physics in terms of two complementary pictures\cite{Susskind}: the in-falling and the external observers. In a nutshell the two descriptions, within our scheme, are roughly as follows. The external observer sees the cascade of quanta of increasing energy that escape and the accompanying change of $N$. Instead, the observer inside the BEC experiences how the number of low-lying Bogoliubov modes diminishes but also how the $b$-operators defining the creation of these low-lying excitations are self similarly transformed along the process. The situation could become paradoxical if we insist in keeping the same operators for the Bogoliubov modes along the process i.e if we do not track the self-similar change of the underlying Bogoliubov transform we have defined in the previous section.}.
 
  The fact that collapse of a black hole is happening in a self-similar way, so that black hole stays at the critical point, can be seen from the energy balance
argument similar to (\ref{deltae}).  For the black hole case   ($g \, =\, \hbar L_P^2$ and  $L_0 \, = \, L$) this energy balance 
  takes the form, 
 \begin{equation}
  E \, \sim  \,  \hbar {N \over L} \,   - \,  \hbar  L_P^2 \,  {N^2 \over L^3} \, .     
 \label{balanceBH}
 \end{equation}
  This fixes the critical value $L\, = \, L_P\sqrt{N}$ and the corresponding 
  energy $E_c \, = \, \sqrt{N}  {\hbar /L_P}$.  Beyond this point the energy balance 
  dictates the system to collapse.  But the collapse happens without moving away 
  from the critical point.  To see this let us estimate the energy gained by the system by shrinking its size by, 
  \begin{equation}
   \Delta L \, = - \,  L_P^2 /L  \, ,
 \label{changeL}
  \end{equation}
 which  corresponds to a change of $\Delta N \, = \, - 1$ provided the
 system stays at the critical point.  The corresponding change of 
 energy is  
  \begin{equation}
  \Delta E \, = \, {\hbar  \over \sqrt{N} L_P} \, 
  \label{changeE}
  \end{equation}
  which is exactly the energy needed to leak out a single  quantum and deplete the system self-similarly, as described by the equation (\ref{deplete}). 
  In fact, as shown in \cite{landau}, the above quantum collapse effectively can be 
  described in the language of a Landau-Ginsburg  Lagrangian for 
  the field $N$,  
   \begin{equation}
  {\mathcal L}_{LG} \, = \,  (\dot{N})^2  \, + \,  {1 \over N} L_P^{-2}  \, +  \, L_P^{-2} {\mathcal  O}(1/N^2) \, .
 \label{LG}
 \end{equation} 
  So the graviton BEC collapses self-similarly due to 
  the fine balance between the depletion and the leakage. 
  Such a balance generically is not exhibited by  other collapsing condensates in which the original $N$  particles get redistributed and become localized within a smaller region of space.  Such condensates can deplete, but they do not 
 necessarily leak at the same rate. This is because in such systems the deep solitonic phase is usually accompanied by a large escape energy  (see, e.g.,  \cite{ueda}). In particular depletion in the solitonic regime is very small and of the same order as depletion in the weak coupling uniform BEC phase.

  Notice, that the condensate may not be purely gravitational but can also include bosons that are subject to other microscopic forces (e.g., gauge forces),  that can counteract the gravitational attraction. In such a case  the depletion of the condensate can be suppressed, and collapse can be prevented. 
  Such a condensate when stabilized at the critical point, describes a quantum 
  portrait of what is classically known as an extremal black hole.

 This phenomenon can be described in the language of the following effective Hamiltonian 
  \begin{equation}
  H\, = \, \hbar L_0  \int \, d^3x \,  |{\bf \nabla}  \Psi|^2 \, - \, \hbar L_P^2 
  \int \, d^3 x \, |\Psi^+\Psi|^2 \,  
  + \, H' \, , 
  \label{hamilton3}
  \end{equation}
where 
\begin{equation}
H' \, \equiv \, \int d^3x \,  {L_P^2 \over \hbar}  \, \left | {\bf \nabla} \int d^3x' {1 \over | x - x'|} 
\, \left (\hbar L_0 \,    |{\bf \nabla}  \Psi|^2 \, - \, \hbar L_P^2 \,   |\Psi^+\Psi|^2 \right ) \right |^2 \, ,
 \label{HN}
 \end{equation}
is the contribution to the energy from the long-range Coulomb-type field produced by the Bose-condensate. 

It is easy to see  how this contribution stabilizes the BEC at the critical point. 
  Without this contribution the system would collapse due to the attractive self-interaction term, which for small $L$  scales as $\sim \hbar L_P^2  N^2/L^3$ 
 and dominates over the first term in (\ref{hamilton3}).   We therefore ignore the latter term in stability analysis.  Now, the same self-interaction energy localized within the region $L$ produces a Coulomb-type field that contributes into the energy through $H'$. This contribution scales  as $E' \, \sim \, \hbar L_p^6 N^4 /L^7$. 
  Correspondingly the $L$-dependence of energy is, 
  \begin{equation}
   E \, \sim \, \hbar L_p^6 N^4 /L^7 \, - \,  \hbar L_P^2  N^2/L^3 \,,  
  \label{totaldelta}
  \end{equation}
  which stabilizes the system at $L \,  = \, \sqrt{N} L_P$. 
 
  \subsection{Black Hole Formation  as Quantum Phase Transition}
  
   The  picture of a black hole as of BEC makes it clear why any scattering 
   process that results into a black hole formation implies {\it classicalization} 
   of gravity \cite{class}.  From what we said above, it follows that any such 
   process can be understood as a formation of graviton BEC and 
   its evolution to the critical point of quantum phase transition.  
    
  In order to fix ideas let us consider a two-particle $|in \rangle $-state characterized by a center of mass energy $E$ and some impact parameter $b$. The gravitational self energy of this system is $E_{gr}\, = \, \frac{E r_g(E)}{b}$, where by $r_g(E)$ we mean the gravitational radius corresponding to energy $E$ i.e $r_g(E)\, = \, E L_P^2$. Let us consider the initial situation with very high total energy $E\, \gg \, \hbar / L_P$ and large impact parameter $b>>r_g(E)$. In these conditions the gravitational self energy $E_{gr}$ is much smaller that the total energy $E$. Irrespectively of this we can describe the gravitational self energy in terms of a gas of $N$ gravitons with occupation number $N\, =\,  \frac{Er_g}{\hbar}$ and typical wave length $L \, = \, b$. We can consider this gas of soft gravitons as a BEC confined in a region of size $b$. What effectively plays the role of the confining potential for these gravitons is the external source, namely the two particles in the $|in \rangle $-state. 
 Obviously, for such a condensate $\alpha N \, \ll \, 1$.   
  Thus, we can assume that this Bose-Einstein condensate is in  {\it weak coupling conditions} in a homogeneous phase. The classical order parameter solving the corresponding Gross-Pitaevskii equation effectively describes the Newtonian weak interaction among the $|in \rangle$-particles. In standard practice this corresponds to the eikonal approximation \footnote{Our aim here is not to enter 
  into the technical subtleties of the trans-Planckian scattering, on which a lot 
  of work has been done since the pioneering papers\cite{tp1,tp2,tp3}. 
  Our aim is to uncover the Bose-Einstein condensate picture of black hole formation in  this scattering, which is the key to understanding classicalization of  UV-gravity \cite{class}.}.  We wish to note that the Bose-Einstein  condensate accounts for the exchange of $N$ gravitons in a ladder.
  
   When we vary $b$ keeping the center of mass energy fixed, what we are doing is changing the coupling $\alpha$ among the gravitons in the Bose-Einstein condensate produced by the center of mass energy.
  To account for this increase in the coupling, using diagrammatic terms, we need to add graviton exchanges among two consecutive rungs of the eikonal ladder. According to our previous discussion we should expect to reach a critical value at which the condensate of gravitons approaches a point of quantum phase transition and becomes self-sustained. This obviously happens when $E$ and $E_{gr}$ are of the same order, or equivalently when $\alpha \, \sim 1/N$ . At this point the system is fully dominated by self-gravity and {\it classicalizes}. 
  
    Again the diagramatic interpretation of this quantum phase transition is quite natural, namely the appearance of a contribution to the imaginary part of the amplitude. The black hole works as a bound state contributing to the imaginary part of the scattering amplitude. The special feature of the quantum phase transition is that the black hole {\it "eats-up"} the Goldstone mode in order to gain entropy. In other words the quantum phase transition of the gravitational BEC unitarizes the ultra-planckian scattering.

  \section{Maximal Packing: Bekenstein and Hawking }
  
   In this section we would like to clarify why the quantum holographic degrees of freedom that we were able to identify in our quantum picture would 
  be impossible to recover in any semi-classical treatment.  The two major
  pillars (as well as mysteries) of  black hole physics are Bekenstein entropy 
  and Hawking radiation.  
   
    Bekenstein tells us that the black hole entropy must scale as the area 
 $S \, \propto \, L^2$.  But, since entropy is dimensionless, the area must be measured in some units. In pure (quantum) gravity the only fundamental parameter of correct dimensionality is the Planck area,  $L_P^2$.  So the 
 entropy must scale as $S \, \propto \, L^2/L_P^2$.  But, $L_P$ is a {\it quantum} 
 length.  So Bekenstein entropy is an intrinsically-quantum entity. 
 Not semi-classical, but quantum.  In particular,  in both classical 
 as well as semi-classical limits,  that are commonly applied to black hole physics,  
  Bekenstein entropy becomes infinite. This is because in both limits, $L_P \, \rightarrow \, 0$. 
  One may find this puzzling, but there is nothing to be scared of. This is exactly how it should be. 
   In fact, this behavior is one of the consistency checks of Bekenstein's entropy
 formula, (see below).    Notice, that one cannot assume that this infinity 
 will be regulated by some cutoff. This is because in gravity the cutoff length is $L_P$ and one should be able to consistently take it to zero.   
 
  In order to understand what is going on, let us focus on the semi-classical black hole limit. This limit is given by, 
 \begin{equation}
  G_N \, \rightarrow \, 0, ~~~ \, L \equiv  M G_N \, = \, {\rm fixed}  \,, ~~~ \hbar \, = \, {\rm fixed} \, , 
  \label{slimit}
  \end{equation}
where $M$ is the black hole mass. In this limit the black hole geometry 
is fixed and one can consider quantum fluctuations on it without worrying 
about the back-reaction.  This is exactly the limit in which Hawking is doing his computation getting an {\it exactly} thermal spectrum of finite temperature 
  $T \, = \, \hbar /R$. But, notice that exactly in the same limit the Bekenstein entropy diverges, since $L_P$ is zero.  
  
   In order to explain why this situation is highly non-trivial and why its clarification requires the 
 physics of Bose-Einstein condensate, we have to put ourselves in 
 the place of a quantum observer.   This observer sees an object of a finite size
 $L$ radiating a thermal spectrum, but having an infinite entropy, or equivalently, 
 an infinite degeneracy of micro-states. 
   If we think of these micro-states being formed by some quantum excitations 
   about the black hole vacuum, we have to admit that each of these 
   infinite number of distinguishable excitations should cost zero energy.  
 How, can an excitation localized within a finite size box cost no energy? 
 Standard quantum mechanical intuition suggest that quantum excitation within the box of size $L$ should cost energy $\sim \, \hbar /L$.  It is true, that a finite size box can possess 
 some zero modes, such as for instance Goldstone zero modes of broken translational invariance,  but usually only a finite number of such modes exist. 
  Thus, where are these infinite number of required zero modes coming from? 
    
     Our quantum picture answers this question in very simple physical terms. 
  The finite size box can house an unlimited number of gapless modes,  because it is a Bose-Einstein condensate with large occupation number $N$ and is at the critical point (\ref{cr}), or equivalently, at the point of maximal packing (\ref{overpack}).   As we have seen, thanks to this criticality, the collective Bogoliubov      
modes cost energy  given by (\ref{gap}), which is $1/\sqrt{N}$-suppressed 
relative to a naive expectation, $\hbar /L$.  This immediately explains 
infinite entropy of the box in the limit $N \, = \, \infty$. 
 
  With the above knowledge everything fits into the place.  
 Hawking's semi-classical limit (\ref{slimit}) in our language is the double-scaling limit (\ref{thelimit}).  In this limit the Bogoliubov energy gap collapses to zero and degeneracy becomes infinite, whereas the Hawking radiation becomes thermal.  
 Correspondingly, the entropy of the black hole becomes infinite. 
 Obviously, working in the semi-classical picture it is fundamentally impossible 
 to trace the origin of holographic Bogoliubov degrees of freedom, since in this 
 limit they decouple as $1/N$, as they should. 
 In other words, any scattering experiment that is aiming to resolve 
 these Bogoliubov modes must have an amplitude suppressed 
 by powers of $1/N$.

   In particular, Hawkings famous "information paradox" is an artifact of the
   semi-classical limit. 
  For $N \, = \, \infty $  Bogoliubov modes can store an infinite amount of information, but it is also infinitely hard to retrieve this information, since 
 modes are decoupled from any observer.  
   But, this is not more surprising than the fact that it is infinitely 
 hard to find a needle in an infinite haystack.

   \section{Quantum Foundation of the Holography}
   
   In our previous paper \cite{Nportrait} we have suggested that the underlying quantum reason for holography was equivalence of the system to large-$N$ 
   BEC.  These left open the question about the true quantum identity of the holographic degrees of freedom. 
   Now we are in a position to suggest a very general answer to this question. 
   We have seen that the graviton condensate that describes a black hole is at the critical point of a quantum phase transition.   The quantum holographic degrees of freedom are degenerate  Bogoliubov modes that become almost gapless at the critical point.  It is striking that the physics at the critical point is described by some sort of CFT. The way black holes manage to store information with a minimal energy cost is through the Bogoliubov quasi zero modes. The holographic bound becomes in this sense a bound on the available number of Bogoliubov zero modes for self sustained condensates.
   
 With the above observations, it is natural to generalize this connection beyond black holes to other gravitational or non-gravitational systems that are expected to have holographic 
description and in our language can be viewed as large-$N$ Bose-Einstein  condensates.  We then suggest that such systems are at the critical point of a quantum phase transition with holographic degrees of freedom being Bogoliubov modes. 
     The first indication that we are on a right track comes in fact from generalizing our reasoning to AdS space. As we have shown, when viewed as a graviton condensate the AdS space also represents a critical point (maximal packing), with the  same relation between the occupation number $N$,  graviton coupling $\alpha$ 
 and the length (radius of AdS) $L$ as in the black hole case.  Remarkably, 
 the occupation number of gravitons coincides with the value of central charge of CFT that has been conjectured \cite{maldacena}  in AdS/CFT correspondence.    Given the fact  the latter conjecture makes no appeal to BEC nature of AdS space,  appearance of the same central charge in our approach  
 is striking. 
  
   Our present picture suggest that this coincidence can have a deep underlying 
   reason.  The  AdS space is a  graviton BEC at the point of critical phase transition.  The appearance of CFT description in such a system is very natural. This is the physics of the corresponding Bogoliubov modes. 
   
   Generalizing the above connection,  we suggest the following
  quantum explanation of holography.  Holographic description can naturally
  arise in non-perturbative  field-theoretic systems, usually described by 
  classical field configurations,  that quantum-mechanically represent large-$N$ BECs at the critical point 
  of quantum phase transition.  The holographic description of such systems 
  is in terms of CFT-type theory of nearly gapless Bogoliubov modes. 
  Notice, that this holographic CFT in $N \, \rightarrow \, \infty$  limit 
  must become lower dimensional. This is because the appearance 
  of the nearly gapless Bogoliubov modes is due to the transition 
  from the uniform condensate to a phase of a bright soliton and is associated 
  with the spontaneous breaking of the translational invariance.  It is natural to expect that gapless Bogoliubov modes must be localized at the edge of the forming bright soliton where gradients of the order parameter become maximal in the solitonic phase.

    Finally, the quantum depletion of BEC should be a measure for a departure 
    from the exact CFT. 
   For example,  the non-extremal black holes deplete and this affects the CFT description by $1/N$-effects.  In contrast,  the AdS and the extremal black holes can be protected from depletion by supersymmetry and extremality and this is probably the reason for a cleaner CFT-description for such systems. 
   This also suggests, why CFT description cannot work for de Sitter  space.   
    
    \section{Outlook}
    
     The purpose of this work was to reformulate a quantum theory of 
 black holes \cite{Nportrait} in the language of condensed matter physics.  
   The key point of the theory is to identify the black hole with a Bose-Einstein condensate 
   of gravitons at the point of {\it maximal packing}. This term refers to a situation when the interaction strength $\alpha$ between the condensed Bosons (gravitons) is equal to the inverse occupation number $N$. 
     It was suggested in the previous work that this property of maximal packing provides a quantum foundation to the known semi-classical properties of black holes. 
 In particular, quantum holographic  degrees of freedom (flavors) responsible for 
 the Bekenstein entropy and information storage, appear as collective nearly gapless excitations of the condensate. 
 
      In the present paper we have shown, that the translation of the above picture in more familiar language of condensed matter systems reveals that physics of the maximally packed graviton condensate is 
 the physics of Bose-Einstein condensates at the critical point of a quantum phase transition, very similar to what have been observed in cold atoms\cite{ueda}.   
 The quantum holographic degrees of freedom are nearly degenerate Bogoliubov degrees of freedom with the mass gap that scales
$\sim \, 1/N$ instead of the inverse size of the system, as one would naively expect.   The magic of  large-$N$ collective effect at the critical point allows to have an unlimited number of nearly-degenerate states within an arbitrarily-small mass gap even for a fixed finite size of the system! 
  
    We have shown, that the black hole graviton BEC remains at the critical point even during 
    the quantum depletion and the collapse.  This quantum collapse of the condensate is nothing but a quantum pre-cursor  of the Hawking radiation. 
    
    It is important to realize that the above holographic degrees of freedom are not reducible to known semi-classical excitations in the 
background black hole metric.   These are Bogoliubov modes of the graviton condensate itself, which are intrinsically-quantum and must decouple as 
$1/N$ in the semi-classical limit (\ref{thelimit}).  

 Our results have a number of interesting implications.  First, they point to 
 a deep underlying connection between the maximally-packed gravitational
(or non-gravitational)  systems and BECs at the critical point of the quantum phase transition. 
 
  This connection offers an intriguing possibility of simulating black hole physics in 
  table-top experiments. 
  
   We have also pointed out that our findings suggest a very general quantum 
   foundation of holography. According to this idea, non-perturbative  
   field configurations that usually are treated classically in reality represent   
 large-$N$ BEC's at the critical point of a quantum phase transition.  These systems 
 admit holographic description in form of  (exact or approximate) CFT of the gapless Bogoliubov  degrees of freedom.  Moreover, the departure from exact 
  CFT must be measured by the quantum depletion properties of the condensate.   
  
   It would be interesting to apply this concept to different systems. 
  The obvious  choices of gravitational systems would be AdS and de Sitter  spaces, 
  which as we have shown in \cite{Nportrait}  obey the 
 large-$N$ BEC properties similar to black holes. 
  
   However,  we have also shown that the ordinary field theoretic topological 
 defects, such as t'Hooft-Polyakov monopole, when viewed as a Bose-Einstein 
 condensate also obey the condition of the maximal packing (\ref{overpack}).            
 Thus, this systems must also be equivalent to BECs at the critical point of a quantum phase transition. The corresponding Bogoliubov modes should then give holographic description of such non-perturbative objects.  This may 
 shed a new useful light at the physics of such objects.

\section*{Acknowledgements}

It is a great pleasure to thank Ignacio Cirac and Peter Zoller for very interesting  discussions on properties of  BEC and some ideas presented here.
It is also a pleasure to thank Philip Walther for discussions on experimental prospects and  Alex Pritzel, Daniel Flassing and Stefan Hofmann   for ongoing discussions on various properties of BEC approach to black holes.
The work of G.D. was supported in part by Humboldt Foundation under Alexander von Humboldt Professorship,  by European Commission  under 
the ERC advanced grant 226371,   by TRR 33 \textquotedblleft The Dark
Universe\textquotedblright\   and  by the NSF grant PHY-0758032. 
The work of C.G. was supported in part by Humboldt Foundation and by Grants: FPA 2009-07908, CPAN (CSD2007-00042) and HEPHACOS P-ESP00346.

\end{document}